# Hydrogen bonds in lead halide perovskites: insights from ab initio molecular dynamics

Alejandro Garrote-Márquez,[a] Lucas Lodeiro,[b] Rahul Suresh,[a,c] Norge Cruz Hernández,[a] Ricardo Grau-Crespo,[d] and Eduardo Menéndez-Proupin[a]

[a] *Departamento de Física Aplicada I, Escuela Politécnica Superior, Universidad de Sevilla, Seville E-41011, Spain*

[b] *Departamento de Química, Facultad de Ciencias, Universidad de Chile, Las Palmeras 3425, Nuñoa 7800003, Santiago, Chile*

[c] *International Research Center of Spectroscopy and Quantum Chemistry - IRC SQC, Siberian Federal University, 79 Svobodny pr., 660041 Krasnoyarsk, Russia*

[d] *Department of Chemistry, University of Reading, Whiteknights, Reading RG6 6DX, UK*

**Abstract**

Hydrogen bonds (HBs) play an important role in the rotational dynamics of organic cations in hybrid organic/inorganic halide perovskites, thus affecting the structural and electronic properties of the perovskites. However, the properties and even the existence of HBs in these perovskites are not well established. In this study, we investigate HBs in perovskites MAPbBr$_3$ (MA$^+$=CH$_3$NH$_3^+$), FAPbI$_3$ (FA$^+$= CH(NH$_2$)$_2^+$), and their solid solution with composition (FAPbI$_3$)$_{7/8}$(MAPbBr$_3$)$_{1/8}$, using ab initio molecular dynamics and electronic structure calculations. We consider HBs donated by X-H fragments (X=N, C) of the organic cations and accepted by the halides (Y=Br, I), and characterize their properties based on pair distribution functions and on a combined distribution function of hydrogen-acceptor distance with donor-hydrogen-acceptor angle. By analyzing these functions, we establish geometrical criteria for HB existence based on hydrogen-acceptor (H—Y) distance and donor-hydrogen-acceptor angle (X—H—Y). The distance condition is defined as $d(\text{H} - \text{Y}) < 3$ Å, for N-H-donated HBs, and $d(\text{H} - \text{Y}) < 4$ Å for C-H-donated HBs. The angular condition is $135° < \sphericalangle(\text{X} - \text{H} - \text{Y}) < 180°$ for both types of HBs. A HB is considered to be formed when both angular and distance conditions are simultaneously satisfied. At the simulated temperature (350 K), the HBs dynamically break and form. We compute time correlation functions of HB existence and HB lifetimes, which range between 0.1 and 0.3 picoseconds at that temperature. The analysis of HB lifetimes indicates that N-H—Br bonds are relatively stronger than N-H—I bonds, while C-H—Y bonds are weaker, with minimal influence from the halide and cation. To evaluate the impact of HBs on vibrational spectra, we present the power spectrum in the region of N-H and C-H stretching modes, comparing them with the normal mode frequencies of isolated cations. We show that the peaks associated with N-H stretching modes in perovskites are redshifted and asymmetrically deformed, while the C-H peaks do not exhibit these effects.

## 1. Introduction

Hybrid organic-inorganic halide perovskites (HOIHP) constitute a versatile family of materials that have attracted much research attention during the last decade. Since first proposed in 2009 by Kojima *et al.*,[1] perovskite solar cells (PSC) have emerged as a strongly promising candidate for efficient and cheap solar cells, and have attained an extraordinary 25.7% record photoconversion efficiency (PCE) as single juntion solar cells, and 32.5% in tandem perovskite/silicon solar cells.[2] Other promising applications of HOIHP include X-ray detectors,[3] LED devices,[4] and water splitting photocatalysts.[5] HOIHP are flexible in terms of morphology and can be sinthetized as bulk crystals, bidimensional laminar crystals, or quantum dots.

The crystal structure of 3D halide perovskites, with general formula $ABY_3$, is well illustrated by methylammonium lead iodide ($CH_3NH_3PbI_3$), represented in **Figure 1a**. The A-site of perovskite is occupied by the organic cation methylammonium ($CH_3NH_3^+$), the B-site is occupied by the lead cation ($Pb^{2+}$, gray balls), and the Y-sites are occupied by iodide anion ($I^-$, violet balls). Methylammonium is often abbreviated as MA or $MA^+$, leading to the short names $MAPbI_3$ or MAPI for the compound $CH_3NH_3PbI_3$. Other members of the family of 3D HOIHP can be generated replacing the MA by another organic cations like formamidinium (FA), $CH(NH_2)_2$ (**Figure 1b** shows the structure of $FAPbI_3$) or by a large inorganic cation like cesium; the iodide anion can be also replaced by bromide or chloride, and the lead cation can replaced by tin or germanium. The full family includes solid solutions of the pure compounds, where each crystal site can be occupied randomly by one or several of the above described components. These mixed compounds show remarkable improvement in their stability (the HOIHP's Achilles heel) and are related with the evolution of record PCE.[6]

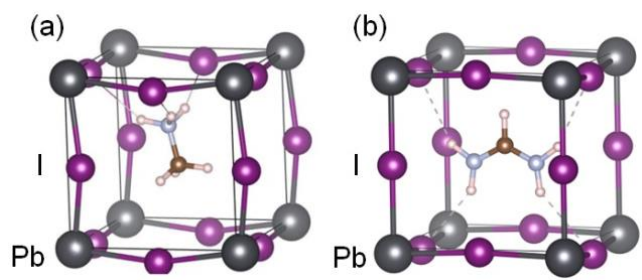

**Figure 1.** Representation of $MAPbI_3$ (a) and $FAPbI_3$ (b) perovskite structures. Hydrogen bonds are indicated by thin dashed lines. Atoms are represented by balls in colors: gray (Pb), violet (I), brown (C), blue (N), white (H). $FAPbI_3$ structure from Ref. 7. Images created with VESTA.[8]

The perovskite structure is generally stable when the ionic radii $r_A, r_B, r_Y$ satisfy an empirical rule $0.8 < t < 1.0$, according to the Goldschmidt tolerance factor:[9]

$$t = \frac{r_A + r_Y}{\sqrt{2}(r_B + r_Y)}.$$

The nature of the A-site cation have a dramatic influence on the phase stability and on crystal symmetry.[10] Substitution of the A-site cation by organic cations larger than FA, with $t > 1.0$, limits crystal growth and allows the formation of laminar structures (2D perovskites),[11] which are periodic in two directions and are a few unit cells wide in the third direction. Another example is $CsPbI_3$, which is stable in the perovskite structure at temperatures just over 600 K, but when cooled to room temperature it makes a fast transition to a yellow non-perovskite phase. A similar behaviour is exhibited by $FAPbI_3$; however, their solid solution, $Cs_xFA_{1-x}PbI_3$, is stable at room temperature.[6] Perovskite phases show a diversity of space groups and phase transitions at different temperatures, depending on the organic cation,[10, 12] as well as on the B-site cation[10] and on the halide.[13]

As **Figure 1** illustrates, the inorganic $PbX_3$ backbone presents large distortions from the ideal perovskite structure. The cells shown in this figure represent formula units of the HOIHP, but do not posses the crystal symmetries inferred from diffraction techniques. Time- and space-dependent realizations of these basic cells conform average structures that make the apparent crystal periodicity in diffraction experiments. Higher- temperature perovskite phases display apparent cubic symmetry, where the crystallographic unit cell coincides with those of **Figure 1**, but that represents only the average positions of the atoms, with large thermal ellipsoids. Moreover, the organic cations rotate and make orientational jumps across all the possible equivalent orientations. For the lower-temperature perovskite phases, the primitive cells are supercells of those represented in **Figure 1**. For example, $MAPbI_3$ undergoes a cubic to tetragonal transition at ~330 K, and the primitive cell has two formula units, and a tetragonal to orthorhombic transition at ~160 K with a primitive cell of four formula units. In the latter case, the cations do not rotate, all atoms keep at definite positions.

Hydrogen bonds (HB) are possible in HOIHP between the halides $X^-$ and the protons bound to nitrogen or carbon atoms, as indicated by dashed lines in **Figure 1**. HB may influence the rotational dynamics of the organic cations, and therefore play a role in the stabilization of the HOIHP, the deformations of the inorganic backbone, and indirectly on the electronic structure.[9, 14-22] The understanding and characterization of HB in perovskites has therefore attracted research attention.

**Figure 2** illustrates the structure of FA and MA. Methylammonium is a result of the protonation of methylamine neutral molecule ($H_2N$—$CH_3$), where the proton is attracted by the lone pair of the $H_2N$ group. The protonated ammonium group ($H_3N$) has three equivalent N—H bonds, therefore the positive charge[23] is considered a property of the full group rather than localized at any of the hydrogen atoms. Hence, the hydrogens of the ammonium extreme are selectively attracted to the halide anions in the perovskites. FA is a result of formamidine ($H_2N$—$CH$=$NH$) protonation. In formamidine ($H_2N$—$CH$=$NH$), the central carbon atom is involved in a single bond with a nitrogen atom and a double bond the other nitrogen atom, leaving a lone pair at latter. However, upon protonation at the NH group, the structure becomes resonant between the Lewis structures $H_2N$—$CH$=$NH_2$ and $H_2N$=$CH$—$NH_2$ . FA has a planar structure, as shown in MD simulations, which can be attributed to the double bond character between the carbon and both nitrogen atoms. The delocalization of the double bond towards the incoming proton can be understood as a positive charge density distributed along the N—C—N backbone, as shown in **Figure 2**. Hence, the attraction exerted by halide anions upon each hydrogen should be smaller than in the MA case due to this distribution of the positive charge. We consider first, the HBs donated by the $NH_2$ groups, but later we will also consider the possibility of donation by the CH group.

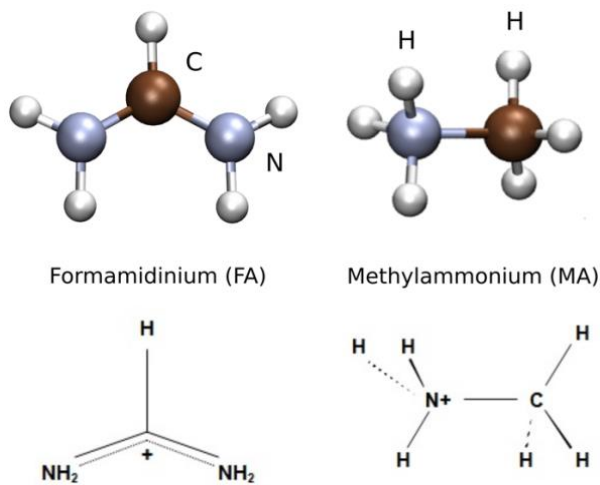

**Figure 2.** Formamidinium (FA) and methylammonium (MA) cation geometry (top) and Lewis structure (bottom). The location of positive charge is indicated, as determined by the extra proton and the electronic cloud distribution.

Svane et al.[24] devised a method to compute the HB energies in hybrid perovskites, filtering out the electrostatic interaction of the charged organic cation with the inorganic lattice. For halide perovskites they obtained energies in the range of 0.02 to 0.27 eV per organic cation, which is one order of magnitude smaller than in formate perovskites. In the high temperature phases of HOIHP, the hydrogen bonds are not permanent but form and break dynamically, allowing cation rotation and determining the symmetry of the crystal. The existence of HB in lead-based HOIHP at room temperature has been recently questioned by Ibaceta-Jaña et al.,[25] based on a combined Raman and density functional theory (DFT) study.

The question of the existence of HB at given conditions require a careful definition of these bonds. According to IUPAC recommendations,[26, 27] HB should display some charge transfer, show directional preference, and show a bond path connecting the hydrogen with the HB acceptor. Early conceptions of HB considered only donors like F-H, O-H and N-H, the last one present in HOIHP. However, the list of donors has expanded to include any molecule or molecular fragment X-H, where X is any element with electronegativity larger than H, i.e., F, N, O, C, P, S, Cl, Se, Br, and I[27]. The acceptor Y can be any of these elements. Therefore the HB donors in HOIHP can be the N-H and C-H groups of the organic cations, while the acceptors can be the halides Cl, Br, and I. Directionality is important criterium, the angle X-H—Y should be close to 180º. Vibrational spectroscopy provides a signature of HBs. The X-H stretching modes typically decrease its frequency (although blue shift is also possible), accompanied by an increase of bandwith and change of intensity in IR spectra. The X-H peak can even dissapear from the spectrum, although this is understood as a large redshift and mixing with other modes.[27] The HBs also modifies the topology of electron density, obtained either from experiments of from electronic structure calculations. In particular, the electron density must present a (3,-1) critical point along the H—Y direction, meaning that the gradient of the density is maximum (minimum) along two (one) orthogonal directions. Methods based on the electron density topology are available to identify HB using quantum chemistry calculations.[21, 28, 29]. Finally, HB need to be thermally stable to have practical significance, but the the absense of effects does not deny its very existence as temporary states.

In this work, we present a systematic theoretical study of the HB features for three HOIHP: $MAPbBr_3$, $FAPbI_3$, and the solid solution $(FAPbI_3)_{7/8}(MAPbBr_3)_{1/8}$, using *ab initio* molecular dynamics (AIMD). For these compounds, we will use the short names MAPBr, FAPI, and MAFA, respectively. We will present statistical functions related to the HBs, such as radial distribution functions, combined (radial and angle) distribution functions, autocorrelation functions and lifetimes, to address some of the open questions about HB in halide perovskites.

**2. Methods**

The $(FAPbI_3)_{7/8}(MAPbBr_3)_{1/8}$ solid solution is represented by a special quasi-random structure (SQS), with an ion distribution that mimics the random distribution in terms of short-range pair correlation functions, as described in previous work.[30] The coordinate files are available in a free repository.[31] Our AIMD simulations at constant volume and temperature (NVT ensemble), with $T= 350$ K, were carried out using the CP2K package.[32] The ionic forces were calculated using DFT in the generalized gradient approximation as implemented in the Perdew–Burke–Ernzerhof (PBE) functional,[33] with the Grimme correction scheme DFT-D3[34] to account for the dispersion interactions. The hybrid Gaussian and plane wave method (GPW)[35] implemented in the QUICKSTEP module of CP2K package was used. The Kohn-Sham orbitals of valence electrons are expanded in a Gaussian basis set (DZVP-MOLOPT for Pb, I, Br, C, N, H).[36] Core electrons were treated using dual-space GTH pseudopotentials.[37-39] Further details of the AIMD calculations can be found in Ref. 30.

The TRAVIS code[40] was used to characterize the HBs via pair distribution functions (PDF), combined distribution functions (CDF), and time correlation functions.[40] The PDF, also called radial distribution function, gives the probability that two atoms of different species are separated a given distance, relative to the probability in a non-interacting gas; the distance of separation is the argument of the function. The PDF reveals the short-range order of a liquid or a disordered material; bond lengths are identified from the peaks at the shortest distances, and coordination spheres are identified from the minima of the PDF. TRAVIS also allows us to obtain other distributions like angle distribution functions and dihedral distribution functions. The CDF are combinations of the former distributions, providing the probability of finding certain combinations of geometrical parameters, such as the pair distance and the angle between three atoms. In particular, the CDF of the halide-hydrogen distance and the halide-hydrogen-nitrogen angle allows to reveal the HBs in HOIHP. The HB dynamics can be characterized by the time correlation functions of the pair forming the HB:

$$C_C^{HB}(\tau) = \frac{1}{N_1 N_2} \sum_{i=1}^{N_1} \sum_{j=2}^{N_2} \int_0^\infty \beta_{ij}(t)\tilde{\beta}_{ij}(t+\tau)dt, \qquad (1)$$

where $\beta_{ij}(t) = 1$ if there is a HB between atoms $i$ and $j$ (one halide and one hydrogen) or zero otherwise. The condition of existence of a HB is defined in terms of a distance and an angle, as will be discussed below. The function $\tilde{\beta}_{ij}(t+\tau) = 1$ if the HB exists for all the time between $t$ and $t+\tau$, and is zero if the HB breaks at any instant. This is called the continuous autocorrelation function in TRAVIS. The function $C_C^{HB}(\tau)$ decays to zero, and the lifetime of the HB can be obtained as:

$$T = 2\int_0^\infty C_C^{HB}(\tau)d\tau \qquad (2)$$

TRAVIS also allows us to compute intermittent autocorrelation functions, where $\tilde{\beta}_{ij}(t+\tau)$ is replaced by $\beta_{ij}(t+\tau)$, which equals unity if the HB exists at $t+\tau$, regardless of whether it breaks or not at intermediate times. The intermittent autocorrelation function is useful for HBs in liquids, but not in HOIHPs because the organic cations are confined in a cavity and any broken HB will form again and again, not allowing the autocorrelation function to decay.

## 3. Results
**HB donated by ammonium and amidine groups**

**Figure 3a** shows the pair distribution functions (PDF) Y—H-N(A) as functions of the halide-hydrogen distance Y—H, in FAPI, MAPBr and MAFA. Here, Y=Br or I, and A=MA or FA, the latter indicating the cation the nitrogen belongs to. The dotted curves represent distribution functions present on the pure compounds, i.e, Br—H-N(MA) in MAPBr, and I—H-N(FA) in FAPI. The corresponding functions for MAFA are shown in solid lines, as well as for the pairs present only in MAFA, i.e., I—H-N(MA) and Br—H-N(FA). The HB is signaled by the sharp peaks between 2 and 3 Å, except for I—H-N(FA) in FAPI (black dotted line) and MAFA (red line). A subset of these functions, those present in pure compounds, were presented in Ref. 30, showing that Y—H-N distances are shorter than Y—H-C distances. The latter case (hydrogen covalently bound to carbon) will be analyzed later. **Figures 3b** and **3c** show that halide-nitrogen distances are smaller than halide-carbon distances, as expected from the positive charge of $NH_3$ and $NH_2$ groups. The effect of shorter Y—N and Y—N-H distances has been reported for MAPI elsewhere.[15, 17]

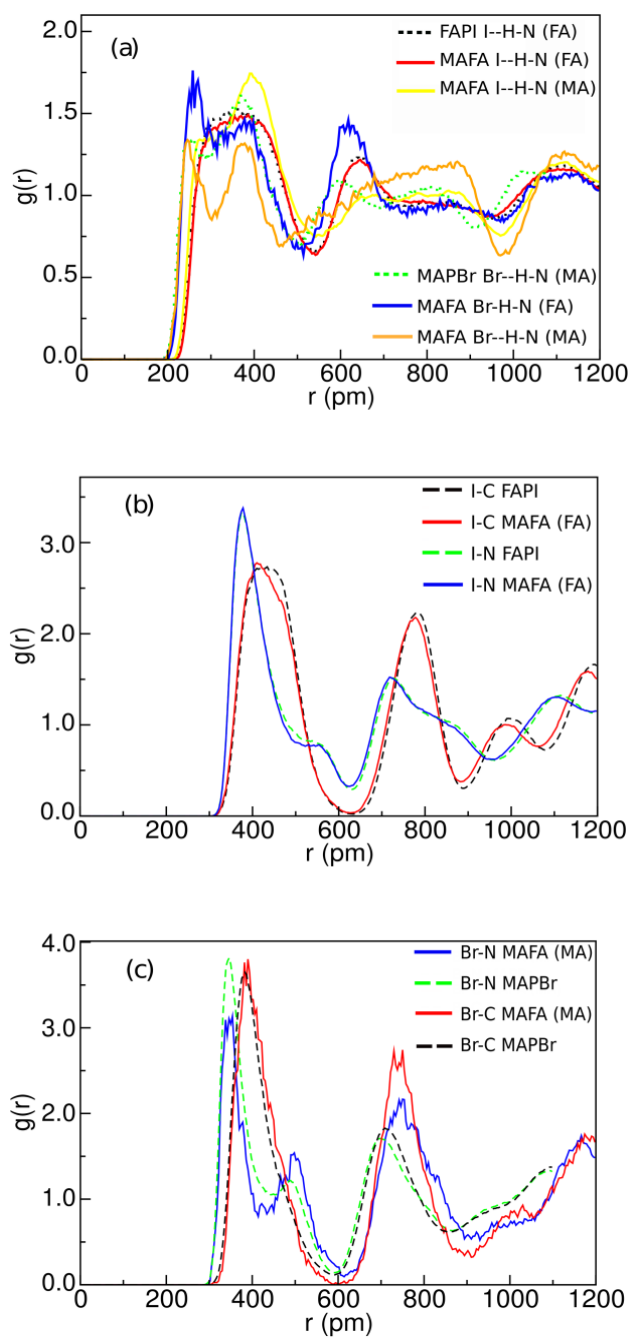

**Figure 3.** Pair distribution functions (PDFs) for a) halide - hydrogen, considering hydrogen covalently bound to nitrogen in MA, FA and MAFA (short name for $FAPbI_3)_{7/8}(MAPbBr_3)_{1/8}$); b) iodide – carbon / nitrogen; and c) bromide – carbon/nitrogen.

HBs are better resolved be means of combined distribution functions (CDF), which are functions of the Y—H distance and the Y—H—N angle. **Figure 4** shows the CDF for the same combinations as in **Figure 3a**. The maximal values take place in the region of the red spots, which can be approximately defined by the simultaneous conditions $d < 3$ Å, and $135º < \sphericalangle(I-H-N) < 180°$. Even for I—H-N(FA) in FAPI and MAFA, which do not show a peak in the PDF, the CDF shows a clear maximum at distances between to 2 and 3 Å followed by minimum between 4 and 5 Å, provided the angle is larger than 135°. Therefore, we take this couple of geometrical conditions as the definition for existence of a HB.

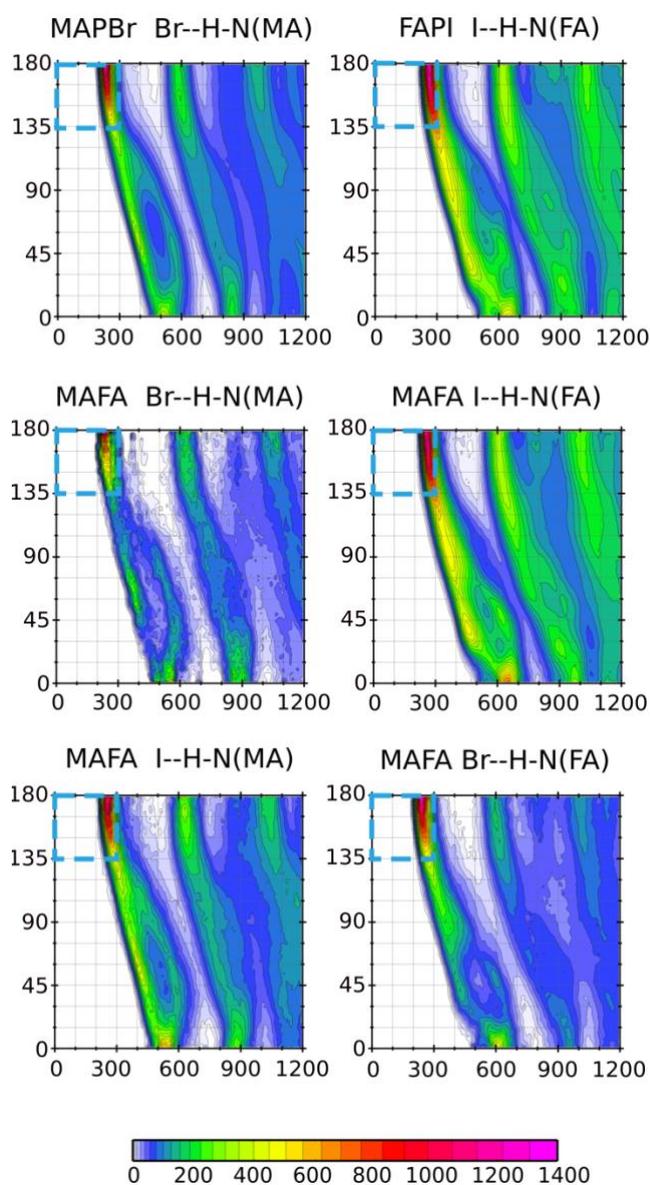

**Figure 4.** Combined distribution functions of Y—H distance (Y=Br, I) in pm (horizontal axes) with $\sphericalangle(Y-H-N)$ angle in degrees (vertical axes) for the pure compounds and the solid solution.

The animation videos V1 to V6 in Supporting Information shows that, with these conditions, HBs form and break dynamically during the simulation time. These animations were generated with the Visual Molecular Dynamics (VMD) graphics program[41], where HBs are defined somewhat differently. HBs in VMD are defined by the Y—N distance instead of the Y—H distance, and the same condition for the angle $\sphericalangle(Y-H-N)$. For these videos, the Y—N distance cutoff was set as 4 Å, *i.e.* 1 Å larger than the Y—H distance cutoff that we have explained above. It is tempting to derive this condition from the CDF of the Y—N distance and Y—N—H angle. This CDF, shown for FAPI and MAPBr in **Figures 5a** and **5c**, shows maxima for distances between 3 Å and 4 Å, but it is almost independent of the angular condition. The snapshots in **Figures 5b** and **5d** shows the Y—N distances smaller than 4 Å (sticks in gray color). These snapshots show that just the distance condition would imply HB with up to three halide atoms simultaneously. However, HBs defined by the combination of distance and angle conditions allow HB bonds with single halides, as shown by the lines in green color. For the HB shown in **Figure 5b**, the distance I—H and the angle I—H-N are 2.68 Å and 154.50°. Similarly, for the HB shown in **Figure 5d**, the distance Br— H and the angle Br—H-N are 3.07 Å and 148.27°.

The animation video V7 in the Supporting Information complements this view, indicating that the first peaks of Y—N PDF seen in **Figures 3b** and **3c** are not explained by HB formation. Proximity of nitrogen and iodine in FAPI has instead steric origin, due to the cation size and the location of N atoms at opposite sites. In the case of MA, the peak of the Br—N PDF also has only a partial contribution from HB. The smaller cation size and the presence of only one N atom suggests that this peak is caused by the electrostatic attraction between the halide and the $NH_3^+$ group and HB. However, the region in the CDF of both **Figures 5a** and **5c**, corresponding to $\sphericalangle(I - H - N) > 135°$, show a depletion for Y—N distance between 5 and 6 Å. This is the signature of HB in this CDF, explaining why the HB can be visualized in VMD using the angular condition together with the Y—N distance. The same behavior can be appreciated for the solid solution MAFA, in **Figure S1** of the Supporting Information.

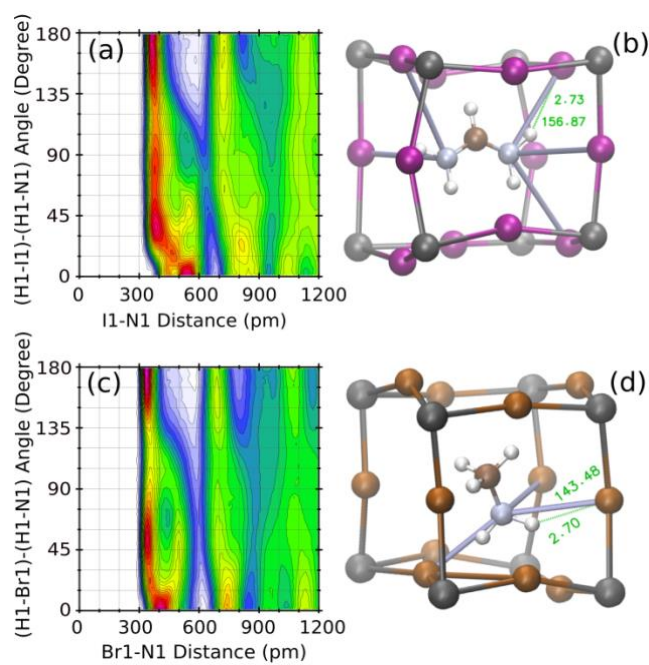

**Figure 5.** Left: combined distribution function of the Y—N distance and Y—H—N angle (Y=I, Br) in FAPI (a) and MAPBr (c). Right: Snapshot of the MD illustrating FAPI (b) and MAPBr (d) with one "real" HB and several fake HB defined by just Y—N distances less than 4 Å.

After the geometrical definition of HB has been established, we now probe their dynamics by means of autocorrelation functions, as defined in Eq. (1), and the lifetimes as defined from Eq. (2). Based on the autocorrelation functions (**Figure 6**), we present the HB lifetimes in **Table 1**. The lifetimes are dependent on the limits set to distance and angle but given a common definition we can observe several trends. One can appreciate that the Br—H-N(MA) and Br—H-N(FA) have the longest duration. Also, it is apparent that the Br—H-N(MA) bond increases its lifetime in the MAFA solid solution (0.27 ps) with respect to MAPBr. However, there are very few Br and few MA in the MAFA model, which leads to poor statistics. The value 0.27 ps is the average of the lifetimes of three symmetrically equivalent hydrogen atoms: two of them have a lifetime of 0.23 ps and the third one has a lifetime of 0.34 ps. The set of all lifetimes, separated by different hydrogens and nitrogens, is shown in **Table S1** of the Supporting Information. HB with iodine has shorter lifetimes than HB with bromine. **Table 1** show HBs with the hydrogens bound to carbon atoms in MA and FA, and also includes the lifetimes of HB with hydrogens attached to carbon atoms, which will be discussed in the next section.

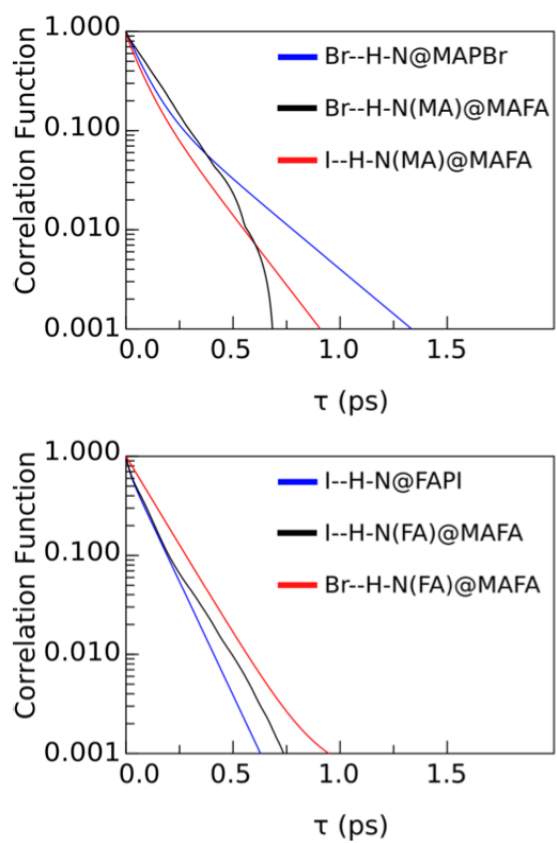

**Figure 6.** Autocorrelation functions of all the Y—H-N bonds studied.

**Table 1.** Lifetimes (in ps) of the different types of HBs derived from the HB continuous time correlation functions.

|  | FAPI | MAPBr | MAFA |
|---|---|---|---|
| I—H-N(MA) |  |  | 0.18 |
| Br—H-N(MA) |  | 0.23 | 0.27 |
| I—H-N(FA) | 0.15 |  | 0.15 |
| Br—H-N(FA) |  |  | 0.23 |
| I—H-C(MA) |  |  | 0.14 |
| Br—H-C(MA) |  | 0.16 | 0.15 |
| I—H-C(FA) | 0.15 |  | 0.16 |
| Br—H-C(FA) |  |  | 0.18 |

**HB through carbon atoms**

We also explore the possibility of HB donated by the C-H groups of MA and FA. Deprotonated FA can be formamidine ($H_2$N-CH=NH) or diaminocarbene ($H_2$N-C-$NH_2$). The latter is highly unstable, but has been found in mass spectroscopy experiments.[42] Regarding HOIHP, several authors[21, 43] have studied the role of Y—H-C bonds, among several non-covalent interactions, and they have pointed importance of HB in determining the extent of the $PbI_6$ octahedra tilting in the low temperature phases. Hence, we proceed to study this possible HB with the same approach used for Y—H-N bonds.

The CDF of the I—H distance with the I—H—C angle, shown for FAPI in **Figure 7a** suggests the HB I—H-C(FA) could be defined from the conditions $d(I-H) < 4$ Å, $90° < \sphericalangle(I-H-C) < 180°$. However, inspection of the MD animation (see video V8) shows that this geometric condition leads frequently to HB of one H with several iodine atoms simultaneously, as shown in Figure 7(c). Considering a more restrictive condition for the angle (**Figure 7a**) like that used to define I—H-N bonds, *i.e.*, $135° < \sphericalangle(I-H-C) < 180°$, a single bond is appreciated, as in **Figure 7b** (also compare supporting video V8 with V1). Let us annotate that the configuration of **Figure 7c** has no HB with the restricted conditions of **Figure 7b**. Moreover, the distance cutoff of 4 Å suggested by the CDF is 1 Å larger than the cutoff for the I—H-N(FA) distance (from the CDF shown in **Figure 4**). Bond distances larger for I—H-C(FA) than for I—H-N(FA), as observed from the location of the red spots in the CDFs of **Figures 4** and **7**, respectively, are usually related with a softer bond. **Figure 7d,e,f** illustrate the corresponding analysis for MAPBr, where **Figure 7e** illustrate the condition used hereafter.

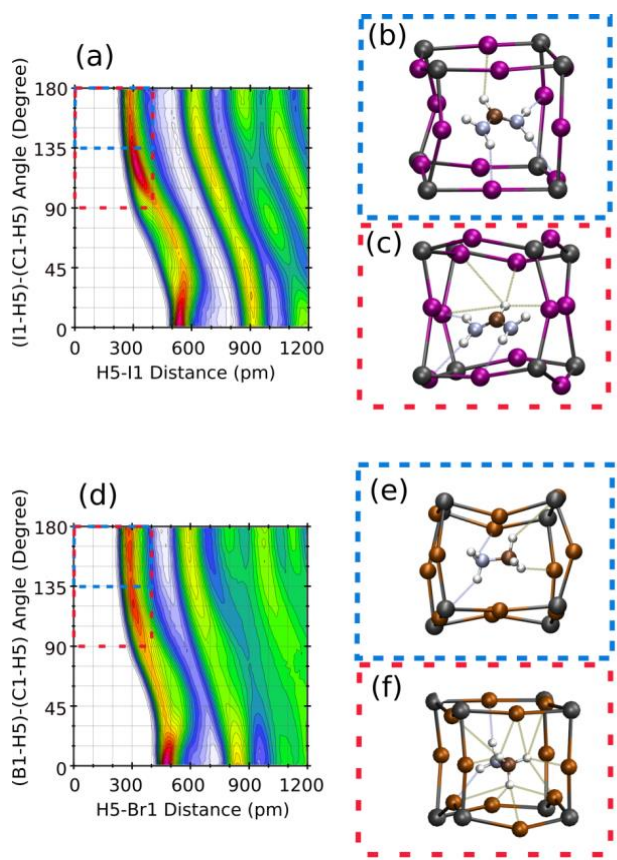

**Figure 7**. Illustration of HB with carbon and nitrogen atoms for FAPI (a-c) and MAPBr (d-f). Images b,c,e, and f are snapshots of the MD simulations, showing the HB by ghostly lines. The CDF (only for Y—H—C angles) for FAPI (a) and MAPBr (d) indicate the range of distances and angles by blue rectangles. The plots at left (right) side indicate a wide (narrow) range of X-H-C and X-H-N angles.

**Table 1** shows that the lifetimes of these HB donated by H-C(FA) are slightly smaller than for the HB donated by H-N(FA). Moreover, there is no difference between Br and I in this respect. The lifetimes of Br—H-C(MA) bonds are close to the former cases, but all of them are smaller than the Br—H-N lifetimes. Naturally, the absolute values of the lifetimes depend on the distance and angle cutoffs that define the existence of the HB. Hence, the comparison of Br—H-C and Br—H-N lifetimes is biased by the difference in the distance cutoffs. Had we set the same cutoff distance for both types of HBs, the Y—H-C HBs would have shorter lifetimes. A broader question is whether these HBs exist at all. The next section presents additional arguments.

**Impact of HB on the power spectrum**

It is well known that HB modify the frequencies of N-H and C-H stretching vibrations,[26, 27] the shape of the power spectrum,[44] providing experimental evidence through IR spectra[45, 46] or Raman spectrum.[20] Hence, we evaluate the associated power spectrum in FAPI and MAPBr. The power spectrum of MAFA has been shown to be well approximated by a weighted average of the spectra of pure compounds.[30] **Figure 8** shows the power spectrum of FAPI (a) and MAPBr (b) in the range 3000-3700 cm$^{-1}$, compared with the vibrational density of states (VDOS) of isolated FA and MA. The frequencies of the normal modes at the minimal energy geometry were obtained by diagonalizing the Hessian of the energy as a function of the internal coordinates. This method provides the power spectrum in the harmonic approximation at zero temperature. The resulting delta-like VDOS has been broadened by means of a Lorentzian function with full width at half maximum (FWHM) of 16 cm$^{-1}$ for each vibrational mode. The peak at 3155 cm$^{-1}$ in **Figure 8a** comes from C-H stretching mode, and the peaks at higher frequencies come from N-H stretching modes. Similarly, for MA, the peak at 3038 cm$^{-1}$ in **Figure 8b** comes from symmetric stretching of C-H bonds in CH$_3$, while the peak at 3152 cm$^{-1}$ comes from two asymmetric stretching modes of C-H bonds. The two peaks at higher frequency also come from N-H stretching modes.

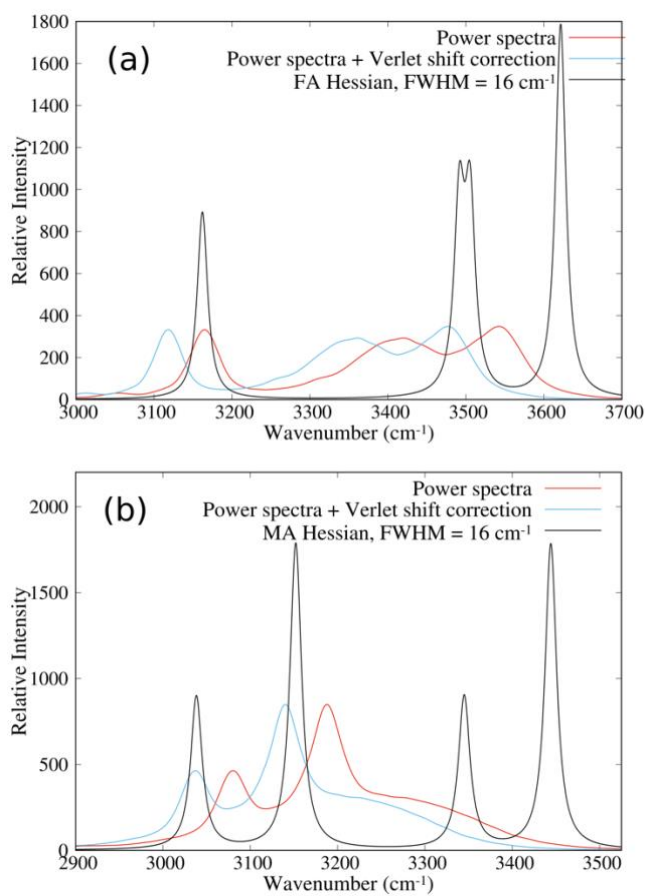

**Figure 8.** Power spectrum of FAPI (a) and MAPBr (b), compared with vibrational DOS of isolated FA and MA obtained from the Hessian matrix.

The power spectra obtained from MD simulations by means of the Fourier transform of the velocity autocorrelation function are shown in red and blue lines for both FAPI and MAPBr. The blue lines include a frequency (wavenumber) correction intended to cancel a systematic error in the oscillator frequencies introduced by the finite time step in the Verlet algorithm.[40, 47] Considering the corrected curves, the vibrational frequencies at the temperature of the MD simulation (350 K), are redshifted from the zero temperature frequencies in all cases. This redshift can be understood as caused by anharmonicity and the environment. Note that peaks of the N-H stretching modes undergo larger redshifts than the C-H stretching mode peaks, but also get broadened asymmetrically toward the low frequency side. This asymmetry of the N-H stretching modes peak is a signature of HB[44]. The peaks of C-H stretching modes do not show asymmetric broadening, as well as none of the other peaks at frequencies under 1800 cm$^{-1}$ (see **Figure S2** in the Supporting Information). This discussion supports the existence of HBs donated by the N-H groups in both FA and MA. It also suggests ruling out the HBs donated by C-H groups.

**Electronic structure**

HBs and other non-covalent interactions (NCI) can be proved in electronic structure calculations by means of the reduced density gradient (RDG)[28]. Varadwaj et al.[21] have shown that the RGD-NCI analysis supports the existence of HB in MAPI. Here, we show that their results also apply to FAPI and MAPBr. The RDG is defined as $s = 1/\left(2(3\pi^2)^{1/3}\right)|\nabla\rho|/\rho^{4/3}$, where $\rho$ is the electron density. As commented above, one of the IUPAC criteria for HB is the existence of a (3,-1) critical point along the X—H path. In a (3,-1) point, the RDG is null, and the Hessian matrix of the density has two negative eigenvalues and one positive eigenvalue, ordered as $\lambda_1 \leq \lambda_2 \leq \lambda_3$. Hence, negative sign of $\lambda_2$ (as for $\lambda_1 \leq \lambda_2$) where $s = 0$ indicates a (3,-1) critical point. Low values of the density indicate non-covalent interactions, in contrast with covalent bonds associated with larger values of the electron density. Hence, there is a certain range of values of the pair $(s, \text{sign}(\lambda_2)\rho)$, that is typical if the vicinity of the (3,-1) critical point of HB. This range is approximately $s \in (0.2, 0.5)$ and $\text{sign}(\lambda_2)\rho \in (-0.05, -0.01)$.[21, 28]

**Figure 9** shows the RDG isosurfaces in unit cells of FAPI (a) and MAPBr (b). The isosurfaces $s(x,y,z) = 0.5$ are shown, using a color scale graded from the function $\text{sign}(\lambda_2)\rho$. The calculations were made using Gaussian09[48] for non-periodic clusters FA$_{19}$Pb$_8$I$_{36}$ and MA$_{19}$Pb$_8$Br$_{36}$, both with charge –1 to have a closed shell electronic structure. The PBE functional was used with the LANL2DZ basis set. The cluster of FAPI was cut from a configuration of the MD ensemble that have one HB minimal I—H(N) distance (2.27 Å), while the cluster of MAPBr was cut from a relaxed periodic structure with a single unit cell, having two HB with Br—H(N) distance equal to 2.43 Å, and $\sphericalangle(\text{Br} - \text{H} - \text{N}) = 168°$. The regions shown in **Figure 9** belong to the central unit cell of each cluster. For both FAPI and MAPBr, we can appreciate a small green cloud ($\text{sign}(\lambda_2)\rho \approx -0.04$) in the middle of each HB (green broken lines), but no similar cloud can be seen near the other hydrogens that do not fulfill the HB geometrical conditions. It is particularly interesting that MAPBr does not show any (3,-1) critical point connecting the third H atom of the ammonium fragment with the nearest Br atom at 2.61 Å. This H atom narrowly misses the HB geometrical criterium because $\sphericalangle(\text{Br} - \text{H} - \text{N}) < 132.5°$.

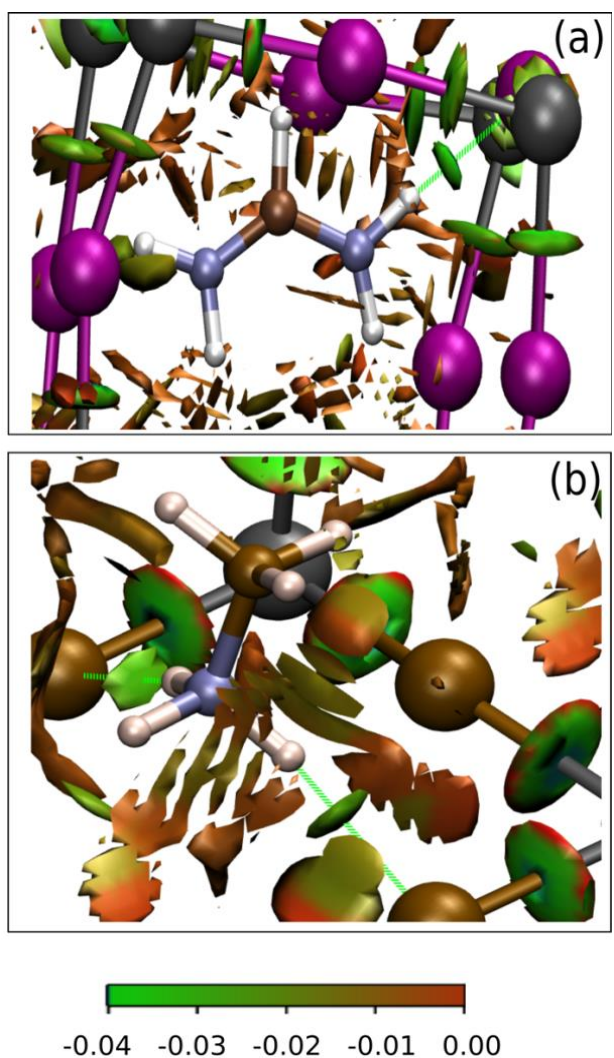

**Figure 9**. RDG plots in FAPI (a) and MAPBr (b), revealing non-covalent interactions. HBs are signaled by green clouds near the centers of the H—I and H—Br paths, appearing only when the geometrical conditions of HBs are fulfilled.

Discussion and conclusions

Let us now discuss some recent publications about HB in HOIHP in the context of our results. Saleh et al[20] have conducted a classical MD study with a huge number of atoms. They concluded that the HBs control the energetics of MA orientations. As they did not consider the directional criterium[26] in their geometrical definition of the HB, they obtained HB populations that are much greater than our estimations. On the opposite side, Ibaceta-Jaña et al.[25] have recently disputed the presence of HB in halide perovskites at room temperature. They proved HBs using Raman spectroscopy, comparing signatures of HB in water, MAI and FAI with the spectra on $MAPbX_3$ and $FAPbX_3$. The absence of spectroscopical evidence does not proof the inexistence of HBs, but rather proofs its irrelevance. However, the failure to observe HBs signatures in their Raman spectra can be attributed to the fact that they centered the analysis on the range 500-1800 cm$^{-1}$, where our simulations also do not show appreciable effects. According to our simulations (**Figure 8**), HBs influence vibrational properties only for the stretching modes N-H over 3000 cm$^{-1}$, agreeing with Ref. 25 in the absence of Raman signatures for lower frequency modes. Moreover, Raman spectra for MAPBr on the range 3000-3500 cm$^{-1}$ reported in the Supporting Information of Ref. 25, seems to agree with our simulations regarding the disappearance of the peaks associated to N-H stretching modes. In any event, the HBs in HOIHP are very weak.

How could we contrast the short HB lifetimes with experimental evidence? Mozur and Neilson[49] have compiled a long list of activation barriers and residence times associated to the movement of organic cations in different HOIHP. Residence times associated to wobble-in-a-cone motion of MA could be related with the HB lifetimes, although an exact correspondence cannot be established. These residence times at room temperature (~0.3 ps) are of the same order of magnitude as our HB lifetimes. These characteristic times show up in rotational time correlation functions obtained from MD simulations.[50] The difference with the lifetime of the HBs can be that the latter can break and form again before a wobbling rotation takes place. To the best of our knowledge, there are no direct measurements of HB lifetimes.

In summary, in this study we have studied the HB in three halide perovskites by analyzing molecular dynamics simulations. These HB has been defined from geometrical criteria, as a function of the distance and angle between the atoms. The presented geometrical criteria are based the angle-distance CDF and the IUPAC recommendations. We have characterized the statistical properties of HB by means of RDF, CDF, and time correlation functions. From the latter we obtained and HB lifetimes, which are smaller for C-H—Y bonds than for N-H—Y bonds (Y=Br, I). The HB lifetimes are similar in the solid solution (FAPbI$_3$)$_{7/8}$(MAPbBr$_3$)$_{1/8}$, and the end compounds FAPbI$_3$ or MAPbBr$_3$.

On the other hand, analyzing the power spectrum, it has been possible to verify that the power spectra peaks related to N-H stretching modes, especially in MA, are modified as expected from the influence of HBs. However, this is not the case for the C-H stretching modes (**Figure 8**). All these suggest that C-H—Y HB are irrelevant for vibrational spectra. The HBs in these compounds are weak, and possibly undetectable at the high temperature that was set in the molecular dynamic simulations. HB should manifest at lower temperature, particularly near the phase transitions. Our simulations were performed at 350 K but this work establishes a theoretical and computational framework for the investigation of hydrogen bonds that can be applied for the analysis of molecular dynamics simulations at other temperatures

**Acknowledgements**

We acknowledge kind support from M. Brehm with the use of TRAVIS. A. L. Montero-Alejo, A. R. Ruiz Salvador and S. Balestra are also acknowledged for useful comments. This work made use of ARCHER2, the UK's national high-performance computing service, via the UK's HPC Materials Chemistry Consortium, which is funded by EPSRC (EP/R029431). We are also grateful to the UK Materials and Molecular Modelling Hub for computational resources in the Young facility, which is partially funded by EPSRC (EP/T022213/1 and EP/W032260/1). Powered@NLHPC: This research was partially supported by the supercomputing infrastructure of the NLHPC (ECM-02).